\documentclass[10pt,compsocconf]{IEEEtran}
\usepackage{algorithm,algorithmic,amsbsy,amsmath,amssymb,epsfig,bbm,mathrsfs,multirow,amsthm}
\usepackage{subfigure,float}

\usepackage[top=0.75in, bottom=0.67in, left=0.67in, right=0.67in]{geometry}

\begin{document}
\title{Cyberattack Detection in Mobile Cloud Computing: A Deep Learning Approach}
\author{Khoi Khac Nguyen$^1$, Dinh Thai Hoang$^2$, Dusit Niyato$^2$, Ping Wang$^2$, Diep Nguyen$^3$, and Eryk Dutkiewicz$^3$ \\
$^1$ School of Information and Communication Technology, Hanoi University of Science and Technology, Vietnam \\
$^2$ School of Computer Science and Engineering, Nanyang Technological University, Singapore \\
$^3$ School of Computing and Communications, University of Technology Sydney, Australia	\vspace{-5mm}	}

\maketitle
\begin{abstract}	
With the rapid growth of mobile applications and cloud computing, mobile cloud computing has attracted great interest from both academia and industry. However, mobile cloud applications are facing security issues such as data integrity, users' confidentiality, and service availability. A preventive approach to such problems is to detect and isolate cyber threats before they can cause serious impacts to the mobile cloud computing system. In this paper, we propose a novel framework that leverages a deep learning approach to detect cyberattacks in mobile cloud environment. Through experimental results, we show that our proposed framework not only recognizes diverse cyberattacks, but also achieves a high accuracy (up to 97.11\%) in detecting the attacks. Furthermore, we present the comparisons with current machine learning-based approaches to demonstrate the effectiveness of our proposed solution. 
\end{abstract}
{\it Keywords-} Cybersecurity, cyberattack, intrusion detection, mobile cloud, and deep learning.

\thispagestyle{empty}
\section{Introduction}

Mobile cloud computing (MCC) is an emerging architecture which has been developed based on the power of cloud computing to serve mobile devices~\cite{Hoang2013MCC}. MCC allows mobile applications to be stored and remotely executed on cloud servers, thereby reducing computing and energy costs for mobile devices. Furthermore, MCC brings a huge profit to cloud service providers by optimizing cloud resource usage through advanced virtualization technologies. Forbes magazine predicts that worldwide spending on public cloud services will grow at a 19.4\% compound annual growth rate (CAGR) from nearly \$70 billion in 2015 to more than \$141 billion in 2019~\cite{ForbesMCC}. However, the MCC is challenged by cybercrime. According to UK Government, 74\% of small firms in the UK experienced a cybersecurity breach, and 90\% of large firms were also targeted in 2014~\cite{2015PWC}.

To counter cyberattacks in MCC, it is crucial to early detect cyber threats, thereby implementing prompt countermeasures to prevent the risks. Currently, there are some approaches proposed to detect and prevent cyberattacks in cloud environment. For example, the authors of~\cite{Cao2015Entropy},~\cite{Ismail2013}, and~\cite{Sahi2017AnEfficient} introduced solutions to detect DoS attacks. Alternatively, one can also rely on attacks' patterns and risk assessment~\cite{Youssef2016Intrusion}, game theory~\cite{Nezarat2017AGame}, and supervised learning~\cite{Nenvani2016Asurvey} to detect and counter cyber threats. The common limitation of these methods is the relatively low accuracy in detecting cyberattacks, and they are unable to work effectively in real-time cloud systems with different types of attacks. In this paper, we propose a framework with an advanced detection mechanism developed from deep learning technique which allows to detect various attacks with high accuracy. 

\begin{figure*}[!t]
	\begin{center}
		\epsfxsize=6.0 in \epsffile{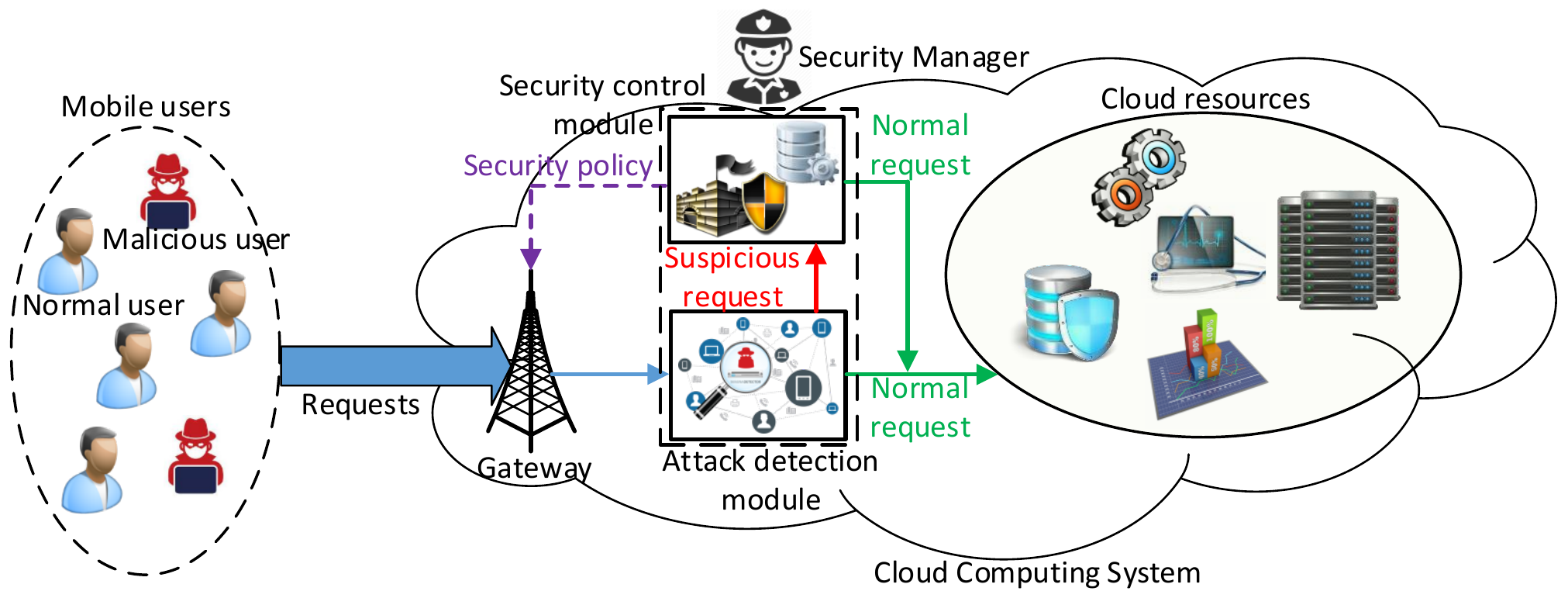}
		\caption{System model for cyberattack detection in mobile cloud computing system.}
		\label{fig:SystemModel}
	\end{center}
\end{figure*}

Deep learning is a sub-field of machine learning concerned with algorithms inspired by the structure and functions of neural networks~\cite{LeCun2015Deep}. Over the past few years, deep learning has been implemented successfully in many areas. For example, deep learning can be used in automatic translation machines to improve the reliability, in recommendation systems to suggest what customers are interested in or want to buy next, and in image recognition systems to detect objects~\cite{LeCun2015Deep}. In this paper, we introduce a framework to detect and prevent cyberattacks through using deep learning technology. The core idea of deep learning method is using a training dataset to train the pre-established neural network in offline mode with the aim to adjust weights of the neural network. Then, the neural network will be used to detect cyberattacks in the cloud system in online mode. Through experimental results, we show that our proposed framework can detect diverse attacks with very high accuracy. In addition, we have also compared the performance of our proposed framework with those of conventional intrusion detection methods to demonstrate the efficiency of our solution. 


\section{Related Work}
\label{sec:RW}

There have been a rich literature dealing with DoS attacks in the cloud computing environment. In particular, the authors in~\cite{Cao2015Entropy} introduced a method taking advantages of virtual machine status including CPU and network usage to identify Denial-of-Service (DoS) attacks in cloud data centers. The authors found that malicious virtual machines exhibit similar status patterns when a DoS attack is launched, and thus information entropy can be applied in monitoring the status of virtual machines to identify attack behaviors. The authors in~\cite{Ismail2013} adopted a convariance matrix method relying on investigating the correlation of several features in the IP header. The authors in~\cite{Sahi2017AnEfficient} developed a classification method to determine the behavior of packets based on Kappa coefficient. 

Besides DoS attacks, other attacks have been also reported and studied. For example, in cloud computing, since multiple Virtual Machines (VM) share the same physical machine, this creates a great opportunity to carry out Cache-based Side Channel Attack (CSCA). A detection technique using Bloom Filter (BF) was developed in~\cite{Chouhan2016Adaptive} to address this problem. The core idea of the BF is to reduce the performance overhead and a mean calculator to predict the cache behavior most probably caused by CSCA. Alternatively, in~\cite{Wang2016Detection}, the SQL injection attack detection method was introduced to prevent unauthorized accesses. The method first obtains SQL keywords through the analysis of lexical regulation for SQL statement, then analyzes the syntax regulation of SQL statement to create the rule tree and traverses ternary tree to detect the attacks. 

The common limitation of existing intrusion detection approaches in cloud environment is that they are unable to simultaneously detect a wide range of attacks with high accuracy. For example, all aforementioned solutions are able to detect one type of attacks only. In this work, we develop an intrusion detection framework which is able to detect diverse attacks in MCC system with high accuracy.  

\section{System Model}
\label{sec:SM}

In this section, we first describe the proposed system model for cyberattack detection along with main functions, and then explain how the system works. As shown in Fig.~\ref{fig:SystemModel}, when a request, i.e., a packet, from a mobile user is sent to the system, it will be passed to the \emph{attack detection module}. This module has three main functions, i.e., data collection and pre-processing, attack recognition, and request processing. 
\begin{itemize}
	\item \emph{Data collection and pre-processing function:} is responsible for collecting data and pre-processing the request to fit the deep learning model. 
	This function is essential to enhance the performance of our model, and helps the gradient descent algorithm used in the training process to converge much faster. 
	\item \emph{Attack detection (online component) function:} is used to classify the incoming requests based on the trained deep learning model. After the deep learning model has been trained in an offline mode, it will be used for this function to detect malicious requests.  
	\item \emph{Request processing:} Given an incoming request, the attack detection function will mark this request as a normal or suspicious request. If the request is normal, it will be served by available cloud resources. Otherwise, the request will be reported to the \emph{security control module}. 
\end{itemize}

When a suspicious request is sent to the security control module, the \emph{request verifying function} will be activated. In particular, the request will be verified carefully by comparing with the current database and/or sending to security service providers for double-checking. If the request is identified to be harmless, it will be served as normal. On the other hand, the request will be treated as a malicious request, and the \emph{attack defend function} will be activated to implement prompt security policies to prevent the spread as well as impacts of this attack. For example, if a request is identified to be a DoS attack, the security manager can immediately implement filters at the gateway to block packets from the same IP address. 
	

\section{Deep Learning Model}
\label{sec:DLM}

In this section, we present the deep learning model for cyberattack detection and explain how the learning model detects cyberattacks in the cloud system. As shown in Fig.~\ref{fig:Deepmodel}, there are two phases in the learning model, i.e., feature analysis and learning process.

\subsection{Features Analysis and Dimension Reduction}

\subsubsection{Features Analysis}

Feature analysis is the first step in the deep learning model. The aim of this step is to extract features and learn from the features. Different types of malicious packets may have special features which are different from the normal ones, and thus by extracting and analyzing the abnormal attributes of packets, we can determine whether a packet is malicious or not. For example, packet features such as \emph{source bytes}, \emph{percentage of packets with errors}, and \emph{IP packet entropy} are important features to detect DoS attacks~\cite{kurihara2014simple}. 

\subsubsection{Dimension Reduction}

Data packets contain many attributes with different features. For example, each record in the KDDcup 1999 dataset~\cite{kddcup1999} and NSL-KDD dataset~\cite{tavallaee2009detailed} consists of 41 features. However, not all the 41 features are useful for intrusion detection. Some features are irrelevant and redundant resulting in a long detection process and degrading the performance. Therefore, selecting features which preserve the most important information of a dataset is essential to reduce the computation complexity and increase the accuracy for the learning process. 

Principal Component Analysis (PCA) is an effective technique which is often used in machine learning to emphasize variation and determine strong patterns in a dataset. The core idea of PCA is to reduce the dimensionality of a dataset consisting of a large number of interrelated variables, while retaining as much as possible the variation presented in the dataset~\cite{jolliffe1986}. Thus, in this paper, we adopt the PCA to reduce dimensions for considered datasets. 

\begin{figure}[!t]
	\begin{center}
		\epsfxsize=3.2 in \epsffile{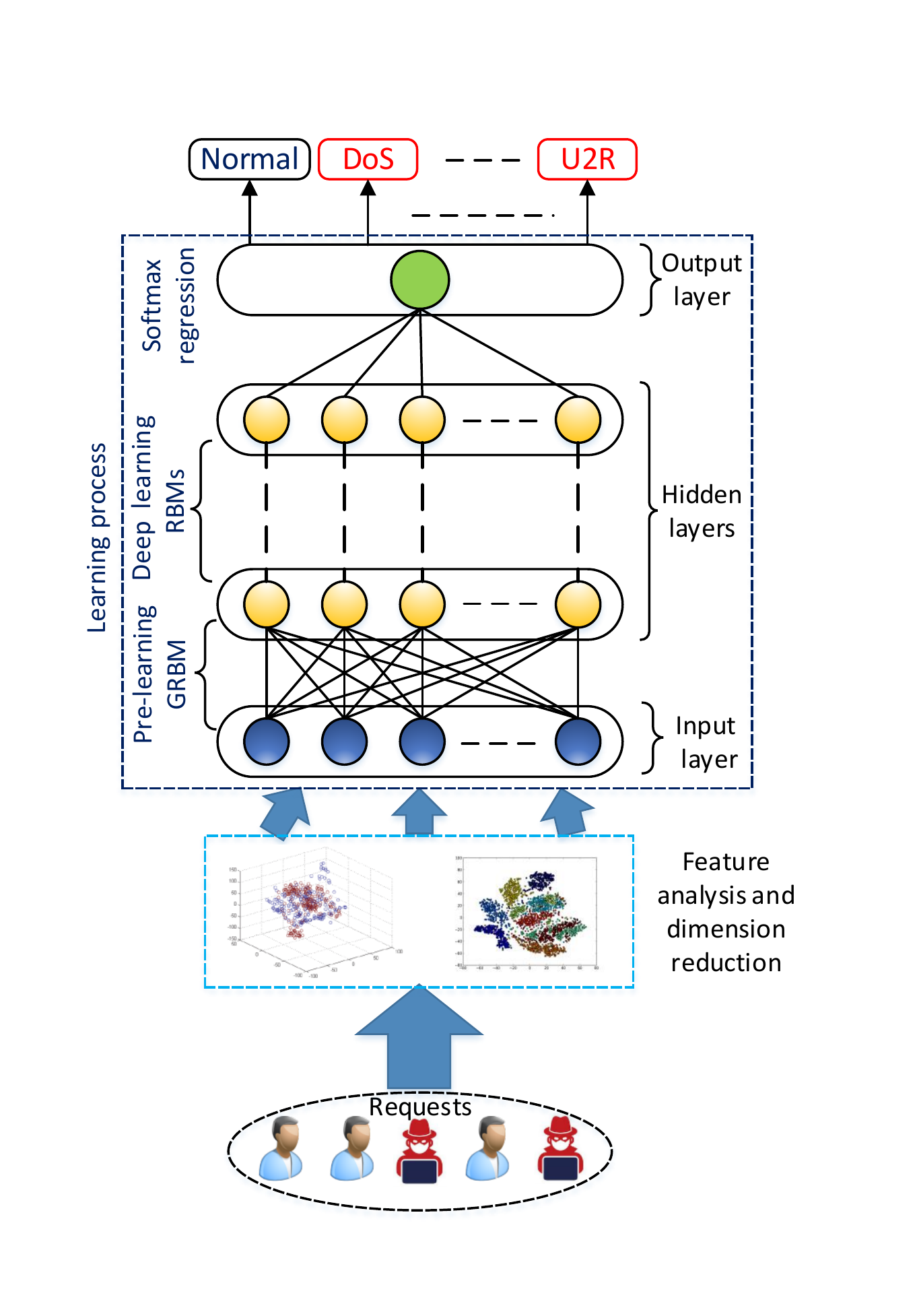}
		\caption{Deep learning model.}
		\label{fig:Deepmodel}
	\end{center}
\end{figure}

Mathematically, the PCA maps a dataset from an $n$-dimensional space to an $r$-dimensional space where $r \leq n$ to minimize the residual sum of squares (RSS) of the projection. This is equivalent to maximize the covariance matrix of the projected dataset~\cite{jolliffe1986}. The dataset in the new domain has two important properties, i.e., the different dimensions of the data have no correlation anymore and the dimensions are ordered according to the importance of their information. We define $\mathbf{X}$ as a $(m \times n)$ matrix with $m$ observations of $n$ different variables. Then, the covariance matrix $\mathbf{C}$ is given by:
\begin{equation}
\mathbf{C} = \frac{1}{n-1} \mathbf{X}^{\top} \mathbf{X} .
\end{equation}

Since $\mathbf{C}$ is a symmetric matrix, it can be diagonalized as follows~\cite{Shlens2014Atutorial}:
\begin{equation}
\mathbf{C} = \mathbf{V} \mathbf{L} \mathbf{V}^{\top} ,
\label{eq:C2}
\end{equation}
where $\mathbf{V}$ is a matrix of eigenvectors and $\mathbf{L} = diag(\lambda _1, ..., \lambda _p)$ is a diagonal matrix of eigenvalues in a decreasing order.
If we use singular value decomposition (SVD) to perform PCA, we will obtain the decomposition as follow~\cite{Shlens2014Atutorial}:
\begin{equation}
\mathbf{X} = \mathbf{U} \Sigma \mathbf{V}^{\top} ,
\label{eq:C3}
\end{equation}
where $\mathbf{U}$ and $\mathbf{V}$ are orthonormal matrices meaning that $\mathbf{U}^T \mathbf{U} = \mathbf{U} \mathbf{U}^T = \mathbf{I}$ and $\mathbf{V}^T \mathbf{V} = \mathbf{V} \mathbf{V}^T = \mathbf{I}$, and $\Sigma = diag(s_1, ..., s_n)$ is a diagonal matrix of singular values $s_i$. Then, we derive the following results:
\begin{equation}
\begin{aligned}
\mathbf{C} =  \frac{1}{n-1} \mathbf{X}^{\top} \mathbf{X}
& = \frac{1}{n-1}(\mathbf{V} \Sigma \mathbf{U}^{\top})(\mathbf{U} \Sigma \mathbf{V}^{\top}) \\
& = \frac{1}{n-1}\mathbf{V} \Sigma^2 \mathbf{V}^{\top} .
\label{eq:C4}
\end{aligned}
\end{equation}

(\ref{eq:C4}) implies that singular vector $\mathbf{V}$ has principal directions and singular value $s_i$ is related to the eigenvalue $\lambda_i$ of covariance matrix $\mathbf{C}$ via $\lambda_i = s_i^2/(n-1)$. Thus, we are able to define the principal components (PCs) as follows~\cite{Shlens2014Atutorial}:
\begin{equation}
\mathbf{P} = \mathbf{X} \mathbf{V} = \mathbf{U} \Sigma \mathbf{V}^T \mathbf{V} = \mathbf{U} \Sigma ,
\label{eq:C5}
\end{equation}
where the columns of matrix $\mathbf{P}$ are the PCs and the matrix $\mathbf{V}$ is called the loading matrix which contains the linear combination coefficients of the variables for each PC. We want to project the dataset from $n$-dimensional to $r$-dimensional while retaining the important dimensions of the data. In other words, we have to find the smallest value $r$ such that the following condition holds:

\begin{equation}
\frac{\sum _{i=1}^{r} \lambda _i}{\sum _{j=1}^{n} \lambda_j} \geq \alpha ,
\label{eq:C6}
\end{equation}
where $\alpha$ is the percentage of information which needs to be reserved after reducing the dimension of input data to $r$-dimensional. We can observe that the PCA will choose the PCs, i.e., important features, that maximize the variance $\alpha$. 

\subsection{Learning Process}

The learning process includes three layers, i.e., input layer, output layer, and some hidden layers, as depicted in Fig.~\ref{fig:Deepmodel}. The refined features will be used as the input data of the input layer. After the learning process, we can determine whether the packet is normal or malicious. The learning process includes three main steps, i.e., pre-learning, deep learning, and softmax regression steps as shown in Fig.~\ref{fig:Deepmodel}.

\subsubsection{Pre-learning Process}

This step uses a Gaussian Binary Restricted Boltzmann Machine (GRBM) to transform real values, i.e., input data of the input layer, into binary codes which will be used in the hidden layers. The GRBM has $\mathcal{I}$ visible units and $\mathcal{J}$ hidden units. The number of visible units (i.e., the number of neurons) is defined as the number of features after reducing dimension, and the number of hidden units is pre-defined in advance. The energy function of the GRBM is defined by~\cite{hinton2010practical}:
\begin{equation}
E(\textbf{v}, \textbf{h}) = \sum^{\mathcal{I}}_{i=1} \frac{(v_i - a_i)^2}{2\sigma_i^2} - \sum^{\mathcal{I}}_{i=1} \sum^{\mathcal{J}}_{j=1} w_{ij}h_j \frac{v_i}{\sigma_i} -  \sum^{\mathcal{J}}_{j=1}b_j h_j ,
\label{eq:C7}
\end{equation}
where $\textbf{v}$ is visible vector and $\textbf{h}$ is hidden vector. $a_i$ and $b_j$ are biases corresponding to visible and hidden units, respectively. $w_{ij}$ is the connecting weight between the visible and hidden units, and $\sigma_i$ is the standard deviation associated with Gaussian visible unit $v_i$. Then, the network assigns a probability to every possible pair of a visible and a hidden vector via the energy function. The probability is defined as follows:
\begin{equation}
p(\textbf{v}, \textbf{h}) = \frac{e^{-E(\textbf{v}, \textbf{h})}} {\sum_{\textbf{v},\textbf{h}} e^{-E(\textbf{v},\textbf{h})}} .
\label{eq:C15}
\end{equation}
From~(\ref{eq:C15}), we can derive the probability that the network is assigned to a visible vector $\textbf{v}$ as follows:
\begin{equation}
p(\textbf{v}) = \frac{ \sum_{\textbf{h}}  e^{-E(\textbf{v}, \textbf{h})}} {\sum_{\textbf{v},\textbf{h}} e^{-E(\textbf{v},\textbf{h})}} .
\label{eq:C18}
\end{equation}
From the probability $p(\textbf{v})$, we can derive the learning update rule for performing stochastic steepest descent in the log probability of the training data as follows:
\begin{equation}
\begin{aligned}
& \frac{\partial \log p(\textbf{v})}{\partial w_{ij}}	= \left< \frac{1}{\sigma_i} v_i h_j \right>_{\mathrm{data}}	- \left< \frac{1}{\sigma_i} v_i h_j \right>_{\mathrm{model}}	,	\\
& \Delta w_{ij} = \epsilon \Bigg(\left< \frac{1}{\sigma_i} v_i h_j \right>_{\mathrm{data}} - \left< \frac{1}{\sigma_i} v_i h_j\right>_{\mathrm{model}} \Bigg) ,
\label{eq:C19}
\end{aligned}
\end{equation}
where $\epsilon$ is the learning rate and $\left< \cdot \right>$ is used to denote the expectation under a distribution specified by the subscript that follows~\cite{hinton2010practical}.

Getting an unbiased sample of $\left< v_i h_j \right>_{model}$ is difficult because there is no connection between hidden units and between visible units in a GRBM. Therefore, sampling methods can be applied to address this problem. In particular, we can start at any random state of the visible units and perform Gibbs sampling alternately. Each iteration of alternating Gibbs sampling involves updating all the hidden units parallelly using~(\ref{eq:C9}) followed by updating all visible units parallelly using~(\ref{eq:C8}).
\begin{eqnarray}
&& p(h_j = 1|\textbf{v}) = sigm \Big(b_j + \sum_i w_{ij} \frac{v_i}{\sigma_i} \Big) ,  \label{eq:C9}	\\
&& p(v_i|\textbf{h}) = \mathcal{N}(v_i|a_i + \sum_j h_j w_{ij},\sigma_i^{2}) , \label{eq:C8}
\end{eqnarray}
where $sigm(x) = 1/(1 + exp(-x))$ is the sigmoid function and $\mathcal{N}(\cdot|\mu,\sigma^{2})$ denotes a Gaussian probability density function with mean $\mu$ and standard deviation $\sigma$. 

\subsubsection{Deep Learning Step}

This step includes a series of learning processes which are performed in sequence to adjust weights of the neural network. Each learning process is performed between two successive layers in the hidden layers through a Restricted Boltzmann Machine (RBM). The RBM is a particular type of Markov random field. It has a two-layer architecture in which the visible binary stochastic units $ \textbf{v} \in \{0, 1\} ^D $ are connected to the hidden binary stochastic units $\textbf{h} \in \{0,1\} ^ F$. Here, $D$ and $F$ are the numbers of visible and hidden units, respectively. Then, the energy of state $\{ \textbf{v}, \textbf{h}\}$ can be calculated by~\cite{hinton2010practical}:
\begin{equation}
E(\textbf{v}, \textbf{h}) = -\sum_{i=1}^{D} \sum_{j=1}^{F} w_{ij}v_ih_j - \sum_{i=1}^{D}a_iv_i - \sum_{j=1}^{F}b_jh_j
\label{eq:C11}
\end{equation}
where parameters $w_{ij}$, $a_i$, and $b_j$ are defined similarly as in~(\ref{eq:C7}).

The conditional probability of a single variable being one (e.g., $p(h_j = 1|\textbf{v})$) can be interpreted as the firing rate of a neuron with the sigmoid activation function as follows~\cite{hinton2010practical}: 
\begin{eqnarray}
&& p(h_j = 1|\textbf{v}) = sigm(\sum_{i=1}^D w_{ij}v_i + b_j) , \label{eq:C13} \\
&& p(v_i = 1|\textbf{h}) = sigm(\sum_{j=1}^{F} w_{ij}h_j + a_i) . \label{eq:C14}
\end{eqnarray}

Similar to the pre-learning step, we can derive the learning update rule for the weights of the RBM as follows:
\begin{equation}
\Delta w_{ij} 
= \epsilon(\left< v_i h_j \right>_{data} - \left< v_i h_j\right>_{model}) ,
\label{eq:C16}
\end{equation}
where $\epsilon$ is the learning rate.

\subsubsection{Softmax Regression Step}

The output of the last hidden layer, i.e., $\mathbf{x}$, will be used as the input of the softmax regression (at the output layer) to classify the packet. A packet can be classified into $M=(K+1)$ classes, where $K$ denotes all types of attacks. Mathematically, the probability that an output prediction $Y$ is class $i$, is determined by: 
\begin{equation}
p(Y = i|\mathbf{x},\mathbf{W},\mathbf{b}) = softmax_i(\mathbf{W} \mathbf{x}+\mathbf{b}) = \frac{e^{W_i \mathbf{x} + b_i}}{\sum_j e^{W_j \mathbf{x} + b_j}} , 
\label{eq:C20}
\end{equation}
where $\mathbf{W}$ is a weight matrix between the last hidden layer and the output layer, and $\mathbf{b}$ is a bias vector. Then, the model's prediction $y_{\mathrm{pd}}$ is the class whose probability is maximal, specifically:
\begin{equation}
y_{\mathrm{pd}} = \arg \max_i \big[ p(Y = i|\mathbf{x},\mathbf{W},\mathbf{b}) \big] 	, \forall i \in  \{1,2,\ldots,M\} 	.
\label{eq:C17}
\end{equation}

\begin{figure*}[!h]
	\begin{center}
		$\begin{array}{ccc} 
		\epsfxsize=2.1 in \epsffile{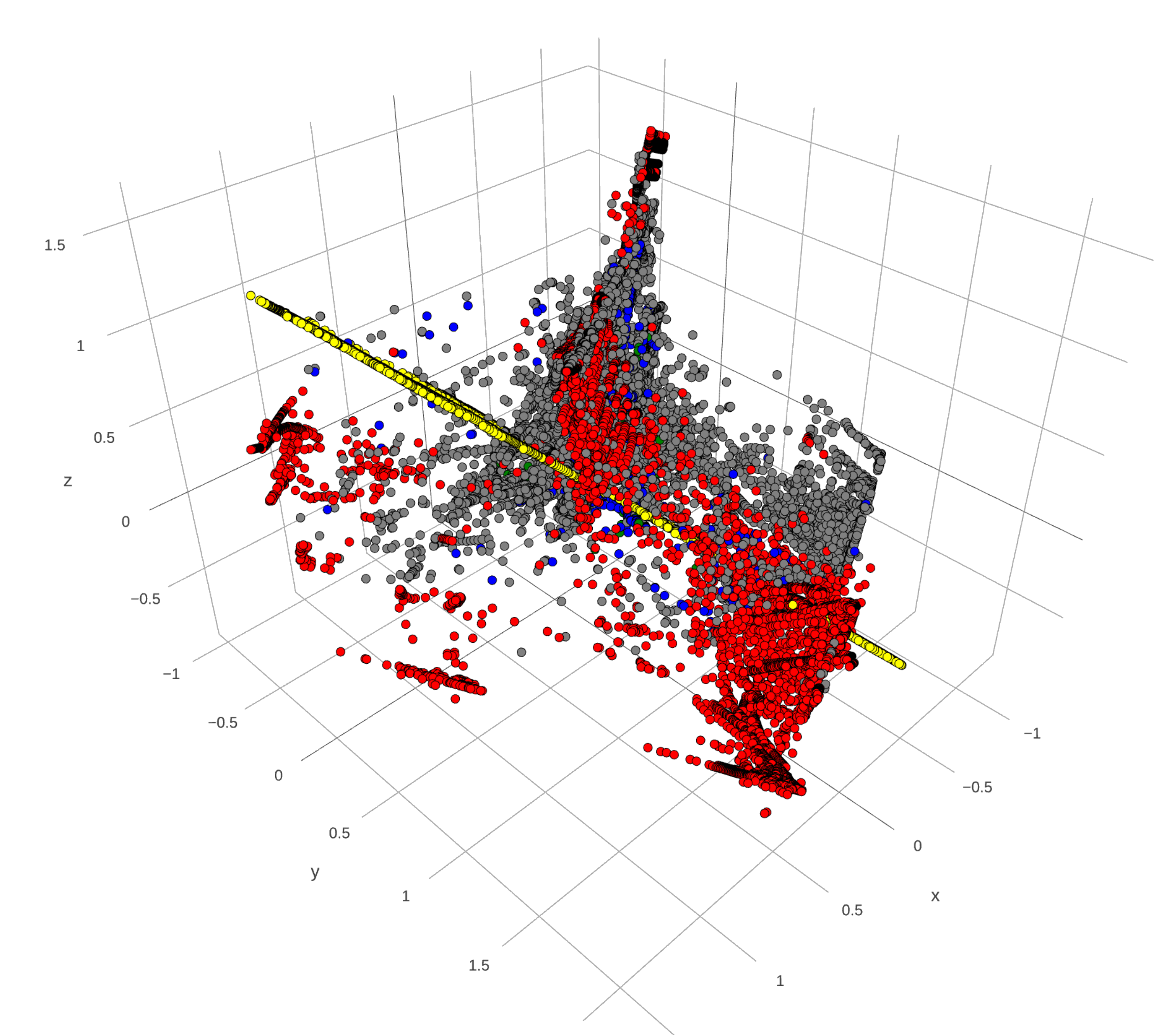} &
		\epsfxsize=2.1 in \epsffile{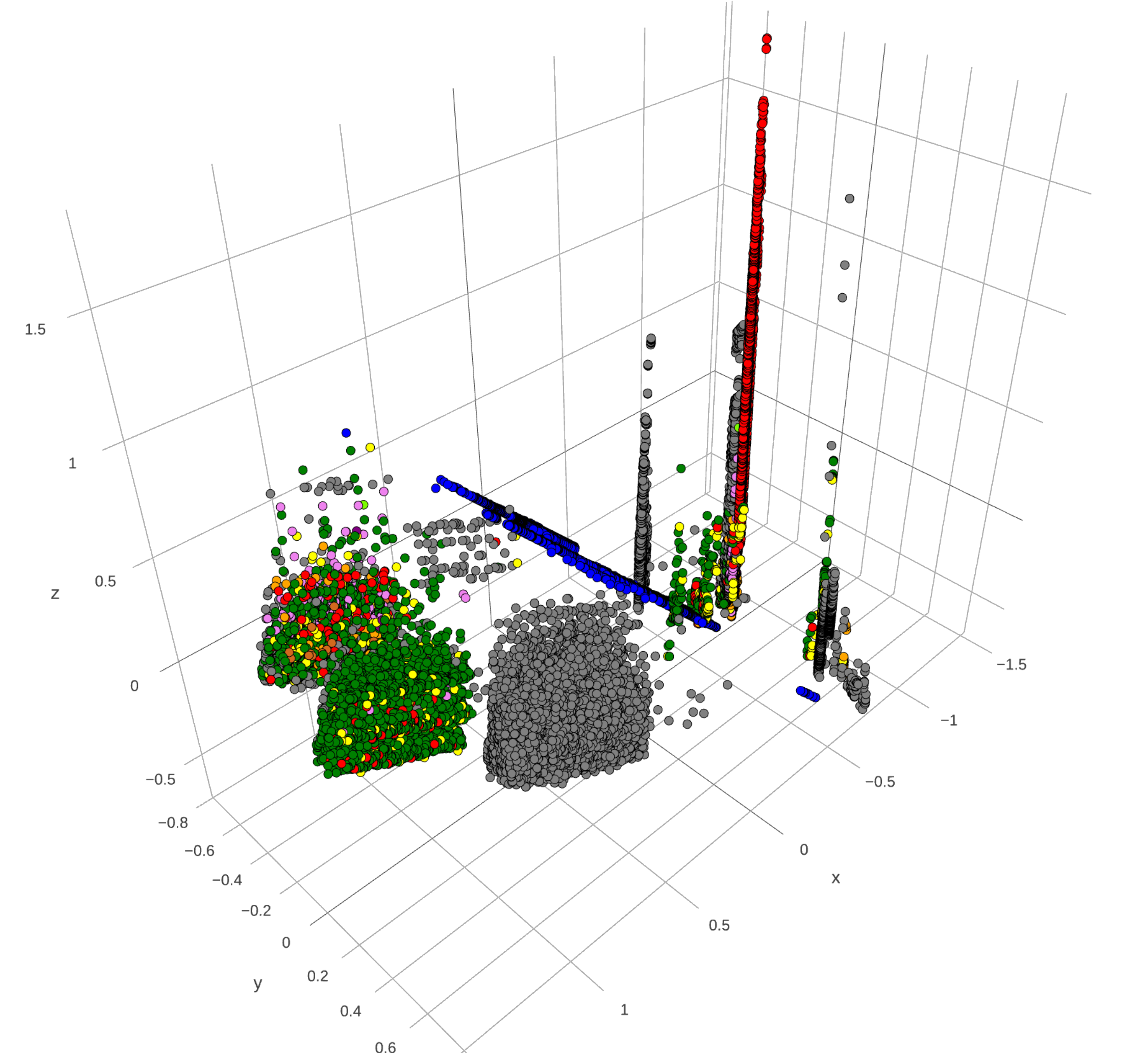} &
		\epsfxsize=2.1 in \epsffile{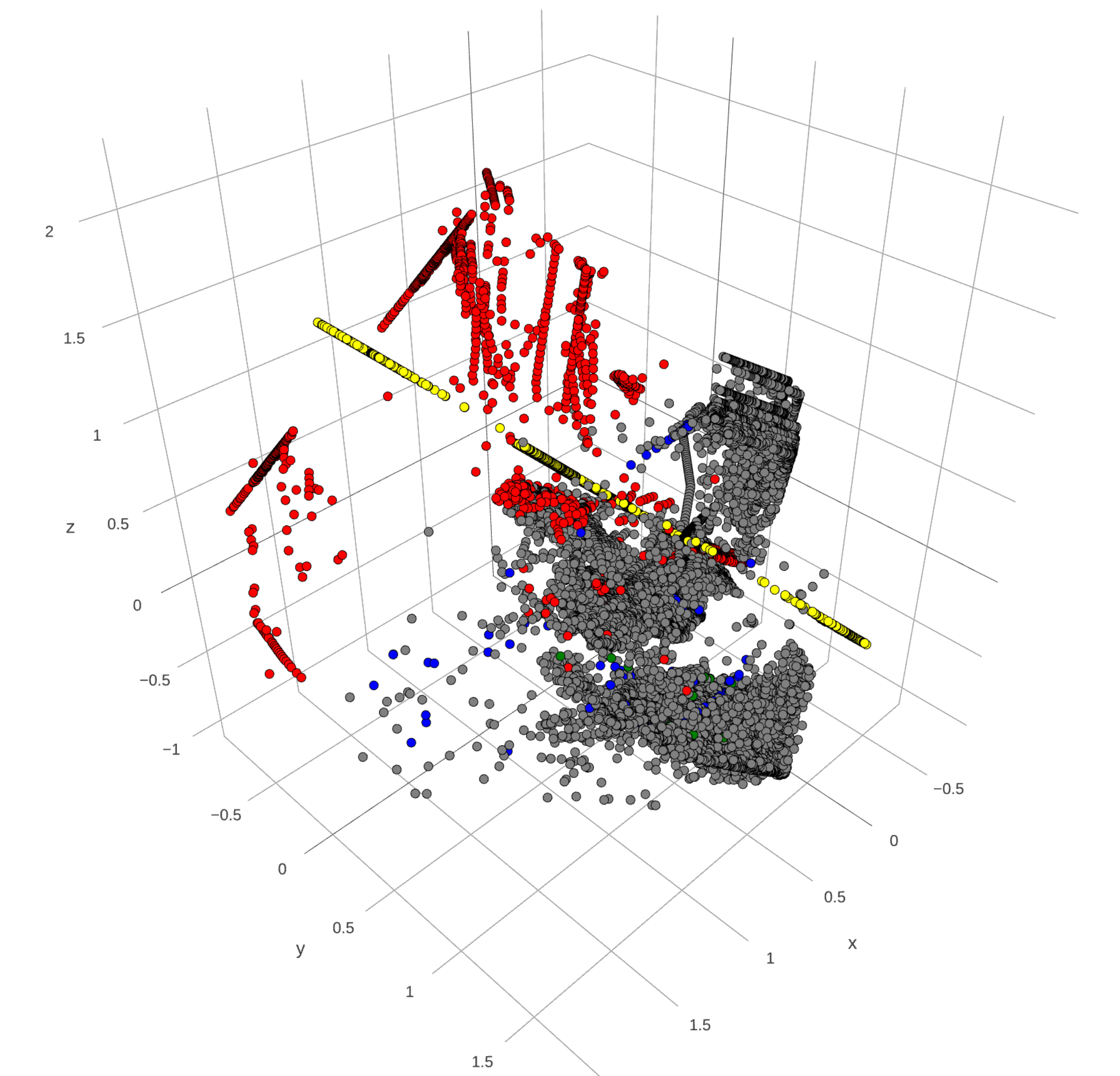} \\ [-0.2cm]
		(a) & (b) & (c)
		\end{array}$
		\caption{Visualizations of three datasets: (a) NSL-KDD, (b) UNSW-NB15, and (c) KDDcup 1999, by using PCA with 3 most important features. Grey circles represent normal packets, while circles with other colors than grey express the different types of attacks.} 
		\label{fig:Visualization}
	\end{center}
\end{figure*}

\subsection{Offline Deep Training and Online Cyberattack Detection}

The deep training consists of two phases, i.e., pre-training phase and fine-tuning phase.
\begin{itemize}
	\item \emph{Pre-training}: This phase requires only unlabeled data which is cheap and easy to collect from the Internet for training. In~\cite{hinton2006fast}, the authors introduced an efficient way to learn a complicated model by using a set of simple sub-models which are learned sequentially. The greedy layer-wise learning algorithm allows each sub-model in the sequence to have a different representation of the data. The sub-model performs a non-linear transformation on its input vectors to produce output vectors that will be used as inputs of the next sub-model in the sequence. The principle of greedy layer-wise unsupervised training for each layer can be applied with RBMs as the building blocks for each layer~\cite{hinton2006fast,hinton2006reducing,bengio2007greedy}. Our training process is executed through Gibbs sampling using CD as the approximation to the gradient~\cite{hinton2002training}.
	
	
	\item \emph{Fine-tuning}: We use the available set of labeled data for fine-tuning. After pre-training phase, we have a sensible set of weights for one layer at a time. Thus, bottom-up back-propagation can be used to fine-tune the model for better discrimination.
\end{itemize}

After the offline deep training is completed, we will obtain a deep learning model with trained weights. This learning model will be then implemented on the attack detection module to detect malicious packets in an online fashion.

\section{Dataset Collection and Evaluation Methods} 
\label{sec:PE}

In this section, we give a brief overview of common cyberattacks in MCC, three real datasets are used in our experiments. We then present the methods to evaluate the experimental results.

%

\subsection{Dataset Collection}

To verify the accuracy of the deep-learning cyber attack detection, we use three empirical public datasets. 

\subsubsection{KDDcup 1999 Dataset}

The KDDcup 1999 dataset~\cite{kddcup1999} is widely used as a benchmark for the intrusion detection network model. Each record in the dataset contains 41 features and is labeled as either normal or a specific type of attack. The training dataset contains 22 types of attacks, while testing dataset contains additional 17 types.

\subsubsection{NSL-KDD Dataset}

The NSL-KDD Dataset was presented in~\cite{tavallaee2009detailed} to solve some inherent problems of the KDDCup 1999 dataset such as the huge number of redundant records both in the training and testing dataset. Each traffic sample has 41 features. Attacks in the dataset are categorized into four categories: DoS, R2L, U2R, and Probe attacks. The training dataset includes 24 attack types, while the testing dataset contains 38 attack types. 

\subsubsection{UNSW-NB15 Dataset}

This dataset has nine families of attacks, namely, Fuzzers, Analysis, Backdoors, DoS, Exploits, Generic, Reconnaissance, Shellcode and Worms. \emph{Argus} and \emph{Bro-IDS} network monitoring tools were used and 12 algorithms were developed to generate totally 49 features. The number of records in the training dataset is 175,340 records and that in testing dataset is 82,331 records from different types of attacks.

\begin{table*}[!t]
	\centering 
	\caption{The comparison between our propose model with other machine learning algorithms on three datasets.}
	\centering 
	\label{table_parameter}
	\centering
	\begin{tabular}{|c||c|c|c||c|c|c||c|c|c||}
		\hline
		&\multicolumn{3}{c|}{NSL-KDD}&\multicolumn{3}{c|}{UNSW-NB15}&\multicolumn{3}{c|}{KDDcup 1999}\\
		\cline{2-10}
		&ACC&PPV&TPR&ACC&PPV&TPR&ACC&PPV&TPR\\
		\hline\hline
		Decision Tree&87.91&63.62&68.50&93.78&76.42&68.92&97.01&94.14&92.52\\ \hline
		K-means&82.78&84.96&56.95&87.05&74.01&35.23&86.19&89.16&65.47\\ \hline
		K Neighbours Classifier&88.56&77.19&71.39&94.31&77.42&71.52&96.85&94.12&92.13
		\\ \hline
		Logistic Regression&89.52&62.04&73.79&92.52&71.05&62.61&96.2&86.29&90.69
		\\ \hline
		Multilayer Perceptron (MLP)&87.91&63.62&68.5&90.16&76.72&75.39&96.77&90.87&91.91
		\\ \hline
		Gaussian Naive Bayes&88.33&73.98&41.67&88.34&73.98&41.67&89.29&83.91&73.22\\ \hline
		Multinomial Naive Bayes&83.96&65.52&59.90&90.97&55.40&54.86&89.66&83.65&74.16\\ \hline
		Bernoulli Naive Bayes&74.60&87.47&36.49&91.31&55.07&56.52&90.94&89.54&77.35
		\\ \hline
		Random Forest Classifier&88.39&71.21&70.99&94.44&80.29&72.21&97.02&94.42&92.56
		\\ \hline
		Support Vector Machine (SVM)&88.32&64.70&70.80&93.38&76.91&66.90&96.74&91.59&91.86\\ \hline
		\textbf{Our Proposed Deep Learning Approach}&\textbf{90.99}&\textbf{81.95}&\textbf{77.48}&
		\textbf{95.84}&\textbf{83.40}&\textbf{79.19}&\textbf{97.11}
		&\textbf{94.43}&\textbf{92.77} \\ \hline
	\end{tabular}
\end{table*}

\subsection{Evaluation Methods} 
\label{subsec:EM}

In this study, we use accuracy, precision, and recall which are typical parameters used in machine learning (deeplearning.net) as performance metrics to evaluate the deep-learning cyberattack detection model. 

\begin{itemize}
	\item \emph{Accuracy (ACC)} indicates the ratio of correct detection over total traffic trace: $\frac{TP + TN}{TP + TN +FP +FN}$
	where $TP$, $TN$, $FP$, and $FN$ stand for ``true positive'', ``true negative'', ``false positive'', and ``false negative'', respectively. Thus, the average prediction accuracy of $M$ supported classes is defined by:  
	\begin{equation}
	ACC = \frac{1}{M} \sum_i^{M}\frac{TP_i + TN_i}{TP_i + TN_i + FP_i + FN_i} .
	\end{equation}
	\item \emph{Precision (PPV)} shows how many attacks predicted are actual attacks. $PPV$ is defined as the ratio of the number of $TP$ records over the number of $TP$ and $FP$ records. 
	
	\begin{equation}
	PPV = \frac{TP}{TP + FP} .
	\end{equation}
	
	\item \emph{Recall (TPR)} shows the percentage of attacks that are correctly predicted versus all attacks happened. $TPR$ is defined as the ratio of number of $TP$ records divided by the number of $TP$ and $FN$ records. 
	\begin{equation}
	TPR = \frac{TP}{TP + FN} .
	\end{equation}
	
\end{itemize}

\section{Experimental Results}
\label{sec:ER}

%

\subsection{Visualizations of Datasets by PCA}

Fig.~\ref{fig:Visualization} illustrates the visualization of three datasets using PCA with $3$ most important features. After using PCA, the normal and malicious packets can be detected effectively with high accuracy. In particular, in Fig.~\ref{fig:Visualization}, normal packets are grouped together and separated from malicious packets. Thus, reducing dimensions in a high dimension dataset not only reduces the computational complexity, but also diminishes significant amount of noise in the dataset, thereby increasing the accuracy in predicting malicious packets.

\subsection{Performance Evaluation}


Table~\ref{table_parameter} compares the performance of the deep learning approach with those of other machine learning algorithms, some of which include K-means, K-neighbors classifier, and the random forest classifier. We observe that the proposed deep learning approach always achieves the best performance in the terms of accuracy, precision, and recall (as defined in Section~\ref{subsec:EM}), for the same datasets. In particular, the accuracy of the deep learning approach is $90.99\%$, $95.84\%$, and $97.11\%$, for NSL-KDD, UNSW-NB15, and KDDcup 1999 datasets, respectively. Furthermore, both precision and recall parameters achieved from the proposed approach are also much higher than those of other machine learning algorithms. 


In summary, there are two important observations from our proposed deep learning model.
\begin{itemize}
	\item \emph{Stability:} Deep learning processes allow to achieve high accuracy (up to 95.84\%) under different settings of layers and the number of neurons per layer. 
	
	
	\item \emph{Robustness and flexibility:} Our deep learning model can be used effectively to detect a variety of attacks on different datasets with very high accuracy. 
\end{itemize}

\section{Summary} 
\label{Sec:Sum}

In this paper, we have introduced a deep learning approach to detect cyber threats in the mobile cloud environment. Through experimental results, we have demonstrated that our proposed learning model can achieve high accuracy in detecting cyber attacks, and outperform other existing machine learning methods. In addition, we have shown the stability, efficiency, flexibility, and robustness of our deep learning model which can be applied to many mobile cloud applications. For the future work, we will implement our proposed deep learning model on the real devices and evaluate the accuracy of the model on the real time basis. Furthermore, the energy consumption and detection time of the deep learning model will be evaluated and compared with other methods.

\bibliographystyle{IEEE}

\end{document}